\documentclass{KapProc} 
\usepackage{epsf}
\normallatexbib
\begin{document}
\newcommand{\eins}{\mbox{$1 \hspace{-1.0mm}  {\bf l}$}}
\newcommand{\ket}[1]{ |  #1  \rangle}
\newcommand{\bra}[1]{ \langle #1   |}
\newcommand{\kett}[1]{ |  #1  \rangle\!\rangle}
\newcommand{\braa}[1]{ \langle \!\langle #1   |}
\newcommand{\scall}[2]{ \langle \!\langle #1   | #2
\rangle\!\rangle}
\newcommand{\scal}[2]{ \langle #1 | #2
\rangle}
\newcommand{\proj}[1]{\ket{#1}\bra{#1}}
\newcommand{\projc}[1]{\ket{#1}_c \,{}_c\bra{#1}}
\newcommand{\proja}[1]{\ket{#1}_a \,{}_a\bra{#1}}
\newcommand{\abs}[1]{ |  #1   |}
\newcommand{\av}[1]{\langle #1\rangle}
\newcommand{\calp}{\mbox{$\cal P$}}
\articletitle{Quantum cloning optimal for \\ joint measurements}
\author{Giacomo Mauro D'Ariano and Massimiliano Federico Sacchi}
\affil{Unit\`a INFM and Dipartimento di Fisica ``Alessandro Volta''\\ 
Universit\`a di Pavia, via A. Bassi 6, I-27100 Pavia, Italy}
\email{dariano@pv.infn.it\ \ \  msacchi@pv.infn.it}

\begin{abstract}
We show that universally covariant cloning is not optimal for achieving 
joint measurements of noncommuting observables with minimum added
noise. For such a purpose a cloning transformation that is covariant
with respect to a restricted transformation group is needed.  
\end{abstract}

\section*{Introduction}
Perfect cloning of unknown quantum 
systems is forbidden by the laws of quantum mechanics \cite{wootters}.
However, a universal optimal cloning has been proposed \cite{bh},
which has been proved to be optimal in terms of fidelity
\cite{oxibm,wer}. A obvious relevant application of such optimal
cloning is eavesdropping in quantum cryptography
\cite{gisin}. However, quantum cloning can be of practical interest
as a tool to engineer novel schemes of quantum measurements, in particular
for joint measurements of noncommuting observables. Quite unexpectedly, 
as we will show in the following, the universally covariant cloning is
not ideal for this purpose. Here, instead, cloning must be optimized
for a reduced covariance group, depending on the desired 
joint measurement, in order to make the measurement over the cloned
copies perfectly equivalent to an optimal joint measurement over the original.  
\par In the following we consider: i) the case of spin 1/2, and the
use of the $1\to 3$ universally covariant cloning to 
achieve a joint measurement of the spin components;  
ii) the case of harmonic oscillator, along with  
a $1\to 2$ cloning transformation 
that is {\em not} universally covariant.  
We show that the POVM obtained in the first case does not lead 
to a minimum-added-noise measurement. On the contrary,  the second 
way allows one to achieve the ideal joint measurement of conjugated variables.
\section*{Universally covariant cloning and joint measurement of spin 
components}
\par The $1\to 3$ cloning map is given by \cite{wer}
\begin{eqnarray}
\rho_3=\frac{1}{2}S_3\left(\ket{\psi}\bra{\psi}\otimes 
\eins^{\otimes 2}\right)S_3\;,
\label{tr-3}
\end{eqnarray}
where $|\psi \rangle $ denotes the input state to be cloned, and 
$S_3$ is the projector on the space spanned by the vectors 
$\{\ket{s_i}\bra{s_i}, i=0\div3\}$,
with $\ket{s_0}=\ket{000}$, $\ket{s_1}=1/\sqrt{3}(\ket{001}+
\ket{010}+\ket{100})$, $\ket{s_2}=1/\sqrt{3}(\ket{011}+
\ket{101}+\ket{110})$ and $\ket{s_3}=\ket{111}$ ($\{\ket{0},\ket{1}\}$ 
being a basis for each spin 1/2 system). 
Independent measurements on the three copies along orthogonal axes
provides the following POVM
\begin{eqnarray}
\Pi(\vec m)&=&
\frac{1}{2}{\mbox{Tr}}_{2,3}
\left[S_3\,\Omega (\vec m)\,S_3\right]\;,
\label{povm}
\end{eqnarray}
where 
\begin{eqnarray}
\Omega (\vec m)=
\frac{1}{8}(\eins + m_x\sigma_x)\otimes (\eins + m_y\sigma_y)\otimes
(\eins + m_z\sigma_z)\;,
\label{pom-3}
\end{eqnarray}
is the product of projectors on the output copies in terms of Pauli
matrices $\sigma _\alpha $, 
and $m_\alpha = \pm 1$ corresponds to each of the outcomes.
Explicit calculation gives 
\begin{eqnarray}
\Pi(\vec m)=\frac{1}{8}\,\left[\eins+\frac{5}{9}\vec m 
\cdot\vec\sigma\right]\;,
\label{povms}
\end{eqnarray}
where the factor $5/9$ comes from the shrinking of the Bloch vector 
due to the cloning map (\ref{tr-3}). In order to have the measurement
unbiased, the spin component outcomes $m_\alpha = \pm 1$ must be
rescaled to $m_\alpha = \pm {9\over5}$. In this way the sum 
of the variances corresponding to the three spin components 
$J_\alpha =\sigma_\alpha /2$ becomes
\begin{eqnarray}
\langle \Delta J^2\rangle=
\frac{1}{4}\left(3 \frac{81}{25}-1\right)=\frac{109}{50}\;.
\label{J2}
\end{eqnarray}
The uncertainty for a measurement performed by projecting onto spin coherent
states \cite{Perelomov} reads \cite{jnt}
\begin{eqnarray}
\langle \Delta J^2\rangle\geq 2j+1\;,
\label{J2c}
\end{eqnarray}
where for $j=1/2$ and pure states the bound is achieved, and is equal
to 2.  So, the joint measurement via universally covariant 
cloning does not achieve the minimum added 
noise.  Notice also that the minimum added noise would be achieved 
by a discrete POVM of the form 
$\Pi(\vec m)=\frac{1}{8}[\eins+\vec m \cdot\vec\sigma]$, with
$m_\alpha = \pm 1$. 
\section*{Cloning for harmonic oscillator and joint
measurement of conjugated variables}
We consider now the $1\to 2$ cloning transformation for a bosonic system 
\begin{eqnarray}
{\cal T}(\varrho)=
\frac 12 P_{c,a}(\sigma )(\varrho \otimes \mathbf 1  _a)
P_{c,a}(\sigma )
\;,\label{trcsig}
\end{eqnarray}
where $\varrho   $ denotes the initial state of the system
to be cloned (in the bosonic mode $c$), mode $a$ supports the second
copy, and  the projector $P_{c,a}(\sigma )$ is given by \cite{jnt}
\begin{eqnarray}
P_{c,a}(\sigma )=S_c(\ln
\sigma )\otimes S_a (\ln
\sigma )\,V (\projc{0}\otimes {\mathbf 1  }_a) V^\dag 
\,S_c^\dag (\ln
\sigma ) \otimes \,S_a^\dag (\ln
\sigma )\;\label{vv2}
\end{eqnarray}
with $|0 \rangle $ denoting the vacuum state, 
$S_d(r)=\exp[r(d^{\dag 2}-d^2)/2]$, and 
$V=\exp [\frac \pi 4 (c^\dag a-c a^\dag )]$. The cloning
transformation in Eq. (\ref{trcsig}) can be realized by a unitary
interaction of modes $c$ and $a$ with an ancillary system, as shown in 
Ref. \cite{cerf}. An experimental realization of this continuous
variable cloning has been proposed in Ref. \cite{dema}, where the
clones correspond to single-mode radiation fields and the cloning
machine is a network of three parametric amplifiers under
suitable gain conditions.  
\par Upon measuring 
the quadrature operators $X_c=(c+c^\dag )/2$ and $Y_a=(a-a^\dag )/2i$ 
over the clones---namely
projecting the output copies on the eigenstates $|x \rangle _c$ and 
$|y\rangle _a$---one
implements the following POVM
\begin{eqnarray}
F_\sigma (x,y)&=&\frac 12 \hbox{Tr} _{a}[P_{c,a}(\sigma )
\projc{x} \otimes \proja{y} P_{c,a}(\sigma )]\nonumber \\&= &
\frac 1\pi 
D_c(x+iy)\,S_c(\ln \sigma )\projc{0}S^\dag _c(\ln \sigma )\,D^\dag _c (x+iy)
\;,\nonumber
\end{eqnarray}
where $D_c(\alpha )=\exp(\alpha c^\dag -\alpha ^* c)$ denotes the
displacement operator for mode $c$. Such kind of POVM is formally a
squeezed-coherent state, and it provides the optimal joint measurement  
of the two noncommuting quadrature operators $X_{\pm \phi}=(c^\dag \,
e^{\pm i\phi}+c\,e^{\mp i\phi})$, 
with $\phi=\hbox{arctg}(\sigma ^2)$. This is shown by the relations 
\begin{eqnarray}
&&\int dx\int dy\, (x\cos \phi \pm y\sin \phi)\,F_\sigma (x,y)=
X_{\pm \phi}\;, \\
&&\int dx\int dy\, (x\cos \phi \pm y\sin \phi)^2\,F_\sigma (x,y)
\nonumber \\&=& X^2_{\pm \phi}+\frac 14 \left|\,\sin(2\phi)\,\right|=
X^2_{\pm \phi}+\frac 12 \left|\,[X_\phi,X_{-\phi}]\,\right|\;,
\end{eqnarray}
namely the outcomes ($x\cos \phi \pm y\sin \phi$) have the same
average values as  the expectations 
of the observables $X_{\pm \phi}$ respectively, 
with minimum added noise.
\par The cloning map in Eq. (\ref{trcsig})  is {\em not} 
universally covariant, but is covariant only under the 
group of  unitary displacement operators, namely 
\begin{eqnarray}
{\cal T}\left (D_c(\alpha )\,\varrho \,D_c^\dag (\alpha )\right)=
D_c(\alpha )\otimes D_a (\alpha )\, {\cal T}(\varrho)\, 
D_c^\dag(\alpha )\otimes  D_a^\dag (\alpha ) 
\;. \end{eqnarray}
\section*{Conclusions}
Measures of quality other than fidelity should be used for
optimality of quantum cloning, 
depending on the final use of the output copies. We have shown that
universally covariant cloning---which maximizes the fidelity---
is not optimal for engineering novel schemes of 
joint measurement of noncommuting observables. 
If one wants to 
use quantum cloning to achieve joint measurements, cloning must be
optimized for a reduced covariance group, as shown here in the case of 
harmonic oscillator.
\par\acknowledgment This work has been supported by the Italian Ministero 
dell'Universit\`a e della Ricerca Scientifica e Tecnologica (MURST)
 under the co-sponsored project 1999 {\em Quantum Information
Transmission And Processing: Quantum Teleportation And Error Correction}. 

\end{document}